# Properties of 1.5 μm synchronous heralded single photon sources based on optical fiber

**Qiang Zhou,[1, 2] Wei Zhang,[1,*] Jie-rong Cheng,[1] Yi-dong Huang,[1] and Jiang-de Peng[1]**

[1] *Tsinghua National Laboratory for Information Science and Technology*
*Department of Electronic Engineering, Tsinghua University, Beijing, 100084, P. R. China*
[2] *q-zhou06@mails.tsinghua.edu.cn*
**Corresponding author: zwei@mail.tsinghua.edu.cn*

A 1.5 μm synchronous heralded single photon source (HSPS) is experimentally demonstrated based on dispersion shifted fiber and commercial fiber components in the paper. Experimental results show that both the preparation efficiency and the conditional second order correlation function $g^2(0)$ increase with the pump light level. Between the two important parameters, tradeoff should be taken in order to obtain a high quality fiber-based synchronous HSPS. A synchronous HSPS with a prepare efficiency of >60% and $g^2(0)<0.06$ is achieved, the multi-photon probability of which is reduced by a factor of more than 16 compared with the Poissonian light sources. The experimental results show a great potential of the fiber-based synchronous HSPS for quantum information applications. 

Reliable single photon sources at 1.5 μm are key elements for quantum information science [1, 2]. As a traditional way, weak laser pulse sources are employed as approximate single photon sources. According to the Poissonian photon statistics property of the laser light [3], an approximate single photon source cannot achieve a high one-photon probability while its multi-photon probability is very low. As a result, the security distance of quantum cryptography system based on the approximate single photon source is limited to be relatively short [4]. An ideal single photon source based on single quantum emitter such as single atom, single quantum dot and single nitrogen-vacancy center in diamond [5], can deterministically emit one and only one photon at a time, allowing for dramatic increase of the security distance of the quantum cryptography system. However, the performance of the ideal single photon source highly depends on the property of material and the photon collection technology. Although optical micro-cavities are employed to enhance the spontaneously emission efficiency of the single quantum emitter utilizing Purcell effect, and to improve the photon collection efficiency through controlling the mode property of the cavity, the total efficiency of such devices needs a further improvement for practical application. Recently, HSPS based on generation of correlated photon pair (CPP) has attracted more and more attentions. In HSPS, one photon of the photon pair is detected, and then an electrical signal is provided as the heralding signal for the appearance of the other photon. The contemporary way to realize HSPS is through spontaneous parametric down-conversion process in bulk crystal, periodically poled crystal waveguide and atomic cascades [6]. However, techniques still need improving for this kind of HSPS, such as collecting the photon into a single-mode fiber, adjusting and stabilizing the experimental setup. More recently, CPPs generated in dispersion shifted fiber (DSF) and micro-structured fiber, with high spectral brightness and single spatial mode over broad wavelength range, have been demonstrated through spontaneous four wave-mixing (SFWM) processes [7, 8]. But the quality of the generated CPP is deteriorated by noise photons from spontaneous Raman scattering (SpRS) processes in these fibers. Two methods are developed to improve the quality of CPP. One is fiber cooling technique through submerging the fiber in liquid nitrogen or liquid helium [9, 10]. The other is using micro-structured fiber with special dispersion property, in which large frequency spanned CPPs are generated [11]. The large frequency spanned CPPs are located at the frequency range far from the first order SpRS, which results in high quality CPPs generation and provides a way to realize good performance HSPSs [12]. However, this scheme highly relies on the dispersion characteristics of the fiber available, requires tunable pump light at 800 nm band or 1.0 μm band and needs optical components operating in a very wide frequency range.

In this Letter, a 1.5 μm synchronous HSPS is realized based on commercial fiber components and DSF with fiber cooling technique. Compared with the scheme of HSPS based on the large frequency spanned CPP generation, our scheme can sufficiently utilize the well developed technology for optical fiber communication system.

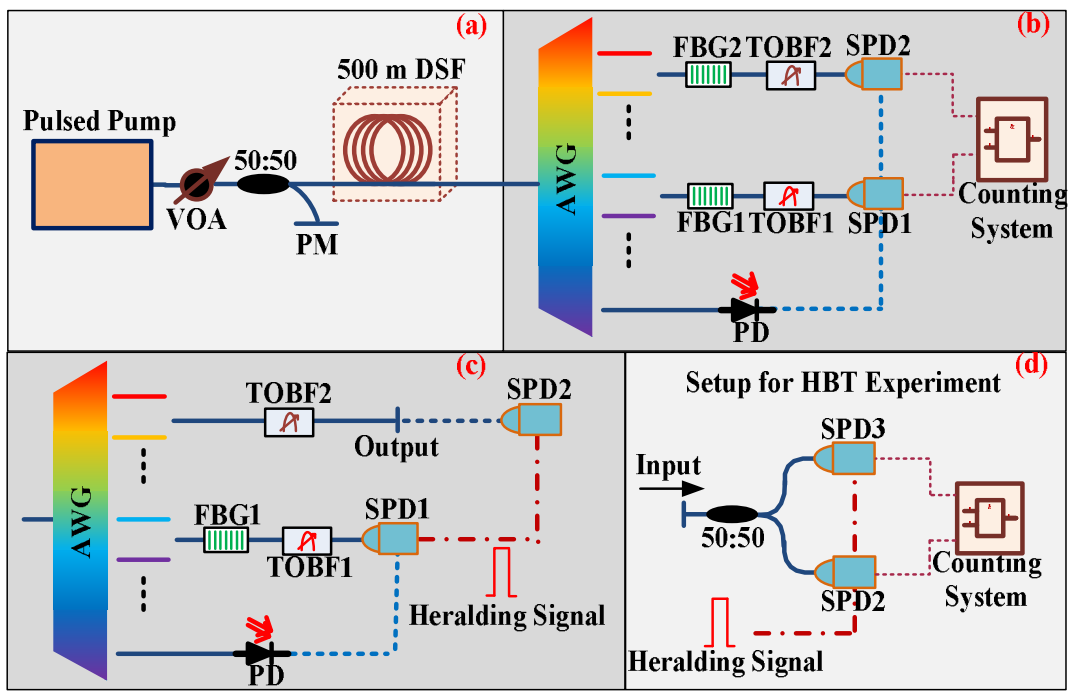

Fig. 1. (Color online) Experiment setup.

The experiment setups for generating CPPs and charactering the synchronous HSPS are shown in Fig. 1. The CPP generation and characterization parts are shown in Fig. 1 (a) and (b), respectively. The pulsed pump is generated by a passive mode-locked fiber laser. The central wavelength of the pump light is 1552.75 nm, with a pulse width of several tens of pico-seconds (estimated by its spectral width) and 1 MHz repetition rate. A side-band rejection of 115 dB for pump light is achieved ensuring the efficient detection of the generated CPPs. A variable optical attenuator (VOA) and a 50/50 fiber coupler with a power meter (PM) are used to control and monitor the pump power level. When pump pulses pass through 500 m DSF ($\lambda_0$=1549 nm, fabricated by Yangtze Co., Ltd.), CPPs are generated through SFWM processes. The DSF is submerged in liquid nitrogen (77 K) to suppress the generation of SpRS photons. The output of the DSF is directed into a filtering and splitting system (in Fig. 1 (b)) based on a 100 GHz/40-channel arrayed waveguide grating (AWG, Scion Photonics Inc.), two fiber Bragg gratings (FBGs), and two tunable optical band-pass filters (TOBFs). Total pump isolation is much greater than 110 dB at either signal or idler wavelength, which guarantees the reliable detection of the generated CPPs. Three InGaAs/InP avalanche photodiode-based single photon detectors (SPDs, Id Quantique, id201) are used to detect the generated CPPs and heralded single photons. All SPDs are operated under Geiger mode with detection window of 2.5 ns and detection efficiency of $\xi_1$ = 16.42%, $\xi_2$ = 21.83% and $\xi_3$ = 22.56% for SPD1, SPD2 and SPD3, respectively. All photon counts are measured with a 30 seconds counting time and averaged after 5 times measurement.

First, the generation of CPPs in DSF submerged in liquid nitrogen is demonstrated and the results are given in Fig. 2. Figure 2 (a) is the results of signal side photon counts. A second order polynomial $N_s=aP_{av}^2+bP_{av}$ is used to fit the experimental data, where $N_s$ is photon count rate of SPD1 triggered by residual pump light detected by a photon-detector (PD), $P_{av}$ is the average power of the pump light, and a = 135.3 and b = 0.8435 are the quadratic and linear fitting coefficients which determine the strengths of SFWM and SpRS processes, respectively. The fitting curve shows that $N_s$ increases quadratically with the pump light level, indicating that SFWM processes dominate the photon generation and the SpRS processes are suppressed by fiber cooling technique. The quantum correlation property of the generated CPPs is demonstrated by observing the ratios between coincidence count ($N_{co}$) and accidental coincidence count ($N_{ac}$) under a fixed idler side wavelength (1550.35 nm) and different signal side wavelengths. Figure 2 (b) shows the experimental results. An averaged ratio of 16.46 is achieved at a signal side wavelength of 1555.15 nm while the averaged ratios are about 1 at other signal side wavelengths, showing the strong quantum correlation property of the generated CPPs.

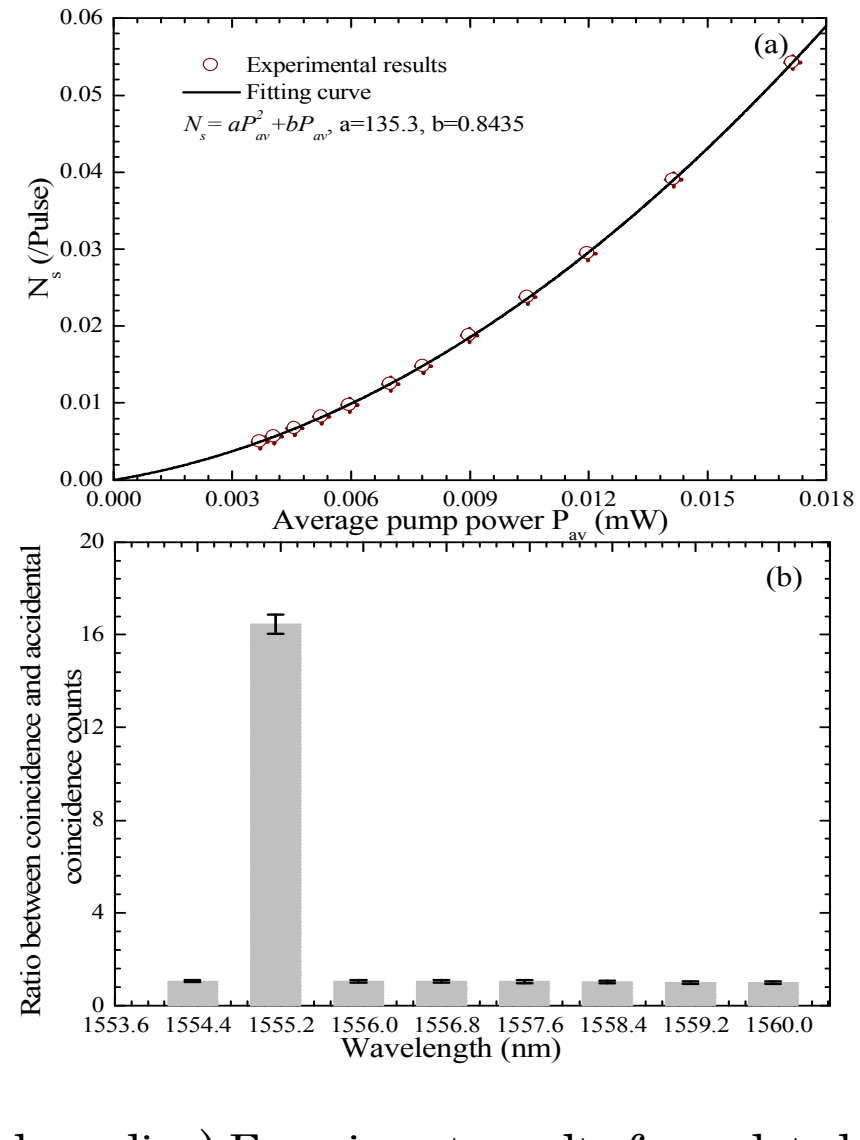

Fig. 2. (Color online) Experiment result of correlated photon pair generation in 500 m DSF.

To realize a synchronous HSPS, the generated CPPs are directed into a modified filtering and splitting system as shown in Fig. 1 (c). In order to improve collection efficiency of the heralded single photon (idler photons in the experiment), the FBG2 on the idler side is removed, since the heralding signal (generated by the detection of the signal photon) can pick the idler photon out, according to the quantum correlation property of the CPPs. In the experiment, output of the SPD1 is used as the trigger for SPD2. The preparation efficiency $\eta_p$, (i.e. probability of an output of idler photon under a heralding signal given by the detection of a signal photon), of the synchronous HSPS is decided by two factors. One is the loss on the heralded photon side of the filtering and splitting system. The other is the noise photon counts on the heralding photon side, such as the SpRS photons, the residual pump photons and the dark counts of SPD1, among which the residual pump photons and the dark counts of SPD1 can be omitted. A loss of $\alpha$ = 2.6 dB on heralded photon side is achieved in the experiment, while a loss of 3.4 dB on heralding photon side. If only considering the loss on the heralded photon side, we can express the theoretical preparation efficiency $\eta_{p0}$ of the synchronous HSPS as

$$\eta_{p0} = 10^{-\frac{\alpha}{10}} \frac{\sum_{n=1}^{+\infty} p(n)\left(1-(1-\xi_2)^n\right)}{\xi_2} \quad (1)$$

where $p(n)$ $(n=1,2,\ldots)$ is the probability of $n$ photons in one heralded photon wave-packet when a heralding signal is given. The solid line in Fig. 3 (a) is the calculated $\eta_{p0}$, the minimum of which is 54.95% determined by the loss on heralded photon side. And due to the increase of the multi-photon probability, $\eta_{p0}$ increases with the photon pair rates.

The practical preparation efficiency $\eta_p$ is obtained by $N_2/(\xi_2 N_1)$, where $N_1$ and $N_2$ are the photon counts of SPD1 and SPD2, respectively. The squares in Fig. 3 (a) are the experimental results of preparation efficiencies under different photon pair rates (the detection efficiency of the SPD2 is divided). It is worth noting that $\eta_p$ is much less than $\eta_{p0}$ while the photon pair rate is small, owing to the influence of the SpRS noise photon. Although the SpRS noise photon can be suppressed by fiber cooling, it is still comparable with the CPPs when the pump light level is very low. On this condition, $\eta_p$ is determined by

$$\eta_p = \eta_{p0} / (1 + N_R / N_c) \quad (2)$$

where $N_R$ and $N_c$ are the photon counts of SpRS and CPP, respectively. Dashed line in Fig. 3 (a) is fitting curve of $\eta_p$ with influence of SpRS noise photon through Eq. (2). The fitting values of $N_R$ under different $N_c$ agrees well with the calculated values obtained in Ref. [13]. It can be seen that $\eta_p$ increases with the photon pair rates and becomes close to $\eta_{p0}$, thanks to the decrease of $N_R/N_c$. And a preparation efficiency of >60% is obtained in the experiment.

Then, $g^2(0)$ of the heralded photons is measured through a Hanbury-Brown and Twiss (HBT) setup (shown in Fig. 1 (d)) [14]. The two output ports of the HBT setup are detected by SPD2 and SPD3, respectively, which are triggered by the output of SPD1. Conditional on the detection of the heralding photon, $g^2(0)$ of the heralded photons is calculated by [15]

$$g^2(0) = \frac{p_{23}(2)}{p_2(1)p_3(1)} \approx \frac{2p(2)}{p^2(1)} \quad (3)$$

where $p_2(1) \approx \xi_2 p(1)/2$ and $p_3(1) \approx \xi_3 p(1)/2$ represent the probability of detecting one photon on the heralded side by detector SPD2 and SPD3, respectively. $p_{23}(2) \approx \xi_2 \xi_3 p(2)/2$ represents the joint probability of simultaneously detecting one photon by SPD2 and SPD3. In the experiment, $g^2(0)$ under different photon pair rates are measured and given in Fig. 3 (b). A $g^2(0)<0.06$ is achieved. It shows that $g^2(0)$ also increases with the photon pair rates due to the increase of the multi-photon probability.

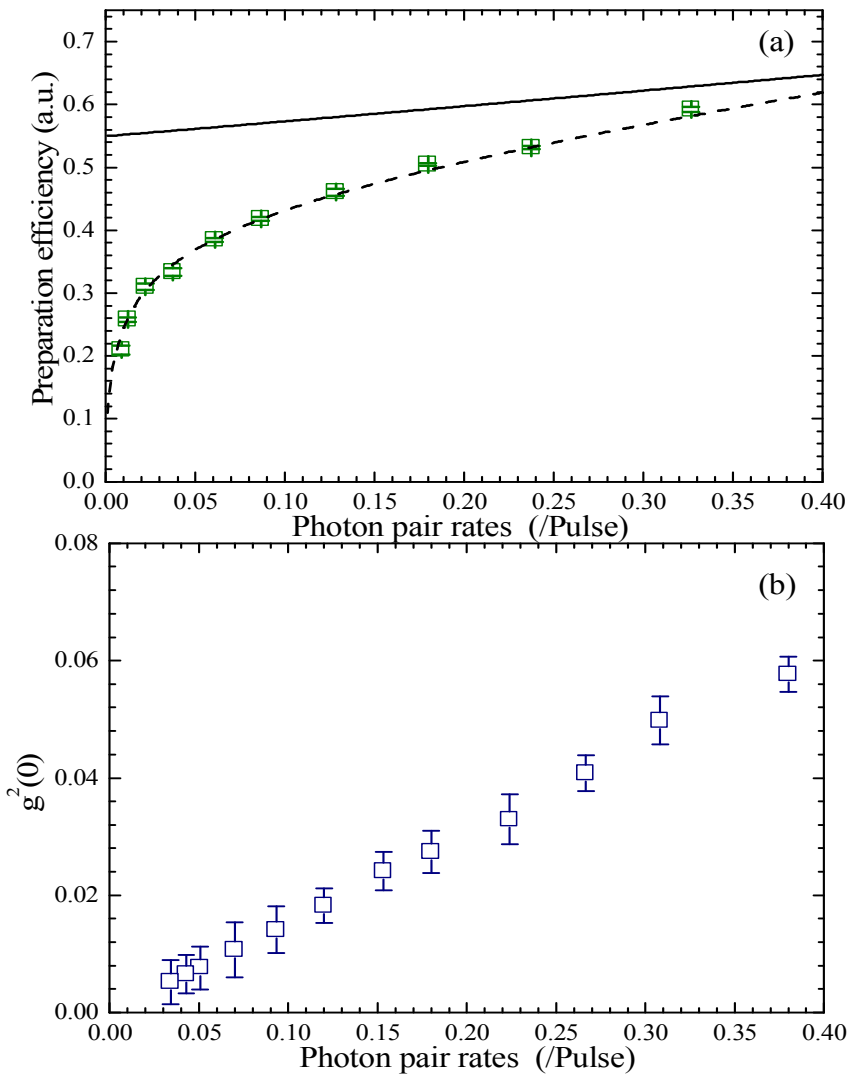

Fig. 3. (Color online) Preparation efficiency and $g^2(0)$ of the synchronous HSPS under different photon pair rates.

For the perfect single photon source and SPDs, $g^2(0)$ equals to 0. For any sources with non-zero probability of multi-photon event, $g^2(0)$ is greater than zero. And a $0<g^2(0)<1$ indicates the non-classical nature of a photon source. In the experiment, $g^2(0)<0.06$ is achieved, which demonstrates the non-classical nature of the synchronous HSPS, while its preparation efficiency is >60%. Compared with the approximate single photon sources, the multi-photon probability of the synchronous HSPS is reduced by a factor of more than 16. Since both the preparation efficiency and $g^2(0)$ increases with the photon pair rate, tradeoff between them should be taken carefully for a high quality fiber-based synchronous HSPS.

In this letter, we have experimentally demonstrated and characterized a 1.5 μm synchronous HSPS based on DSF and commercial fiber components. Experimental results show that both its preparation efficiency and $g^2(0)$ increase with the pump light level. In our experiment, a synchronous HSPS with a preparation efficiency of >60% and $g^2(0)<0.06$ is achieved, which shows the great potential of the fiber-based synchronous HSPS for quantum information applications.

This work is supported in part by National Natural Science Foundation of China under Grant No. 60777032, 973 Programs of China under Contract No. 2010CB327600, Science Foundation of Beijing under Grant No. 4102028, and Basic Research Foundation of Tsinghua National Laboratory for Information Science and Technology (TNList).